\documentclass{article}
\usepackage[T1]{fontenc} 
\usepackage[utf8]{inputenc} 
\usepackage{ismir,amsmath,cite,url}
\usepackage{graphicx}
\usepackage{color}

\usepackage{multirow}
\usepackage{afterpage}
\usepackage{amsmath}
\usepackage{physics}
\usepackage{amsfonts}
\usepackage{amssymb}
\usepackage{graphicx}
\usepackage{comment}
\usepackage[T1]{fontenc}
\usepackage{soul}
\usepackage{afterpage}
\usepackage{booktabs}
\usepackage[parfill]{parskip}
\usepackage[usenames,dvipsnames,svgnames,table]{xcolor}
\usepackage{subfig}

\usepackage{arydshln}
\usepackage{stackengine}
\usepackage{accents}
\usepackage{algorithm}
\usepackage{algpseudocode}

\newcommand\obar[1]{\accentset{\rule{.8ex}{.01ex}}{#1}}
\newcommand\ubar[1]{\stackunder[1.2pt]{$#1$}{\rule{.8ex}{.01ex}}}

\graphicspath{{images/}}

\usepackage{mathrsfs}
\usepackage{lineno}

\usepackage{musicography}

\usepackage{xurl}

\patchcmd{\thebibliography}
  {\settowidth}
  {\setlength{\parsep}{1.7pt}\setlength{\itemsep}{1.7pt plus 1.8pt}\settowidth}
  {}{}

\title{Automatic music mixing with deep learning and out-of-domain data}

\multauthor
{Marco A. Mart\'{i}nez-Ram\'{i}rez$^{\musFlat{}}$ \quad Wei-Hsiang Liao$^{\musFlat{}}$ \quad Giorgio Fabbro$^{\musSharp{}}$ \quad Stefan Uhlich$^{\musSharp{}}$} {\bfseries{Chihiro Nagashima$^{\musFlat{}}$ \quad Yuki Mitsufuji$^{\musFlat{}}$}\\
 $^{\musFlat{}}$Sony Group Corporation, Tokyo, Japan \quad $^{\musSharp{}}$Sony Europe B.V., Stuttgart, Germany\\
{\tt\small marco.martinez@sony.com, yuhki.mitsufuji@sony.com}
}

\def\authorname{M. Mart\'{i}nez-Ram\'{i}rez, W. Liao, G. Fabbro, S. Uhlich, C. Nagashima, and Y. Mitsufuji}

\begin{document}

\maketitle

\begin{abstract}
Music mixing traditionally involves recording instruments in the form of clean, individual tracks and blending them into a final mixture using audio effects and expert knowledge (e.g., a mixing engineer). The automation of music production tasks has become an emerging field in recent years, where rule-based methods and machine learning approaches have been explored. Nevertheless, the lack of dry or clean instrument recordings limits the performance of such models, which is still far from professional human-made mixes. We explore whether we can use out-of-domain data such as wet or processed multitrack music recordings and repurpose it to train supervised deep learning models that can bridge the current gap in automatic mixing quality. To achieve this we propose a novel data preprocessing method that allows the models to perform automatic music mixing. We also redesigned a listening test method for evaluating music mixing systems. We validate our results through such subjective tests using highly experienced mixing engineers as participants.

\end{abstract}
\section{Introduction}
\label{sec:introduction}

\begin{figure}[!t]
\centering
\vspace{-3.25ex}
\centerline{\includegraphics[trim=0cm 0.5cm 0cm 0cm, clip, width=\columnwidth]{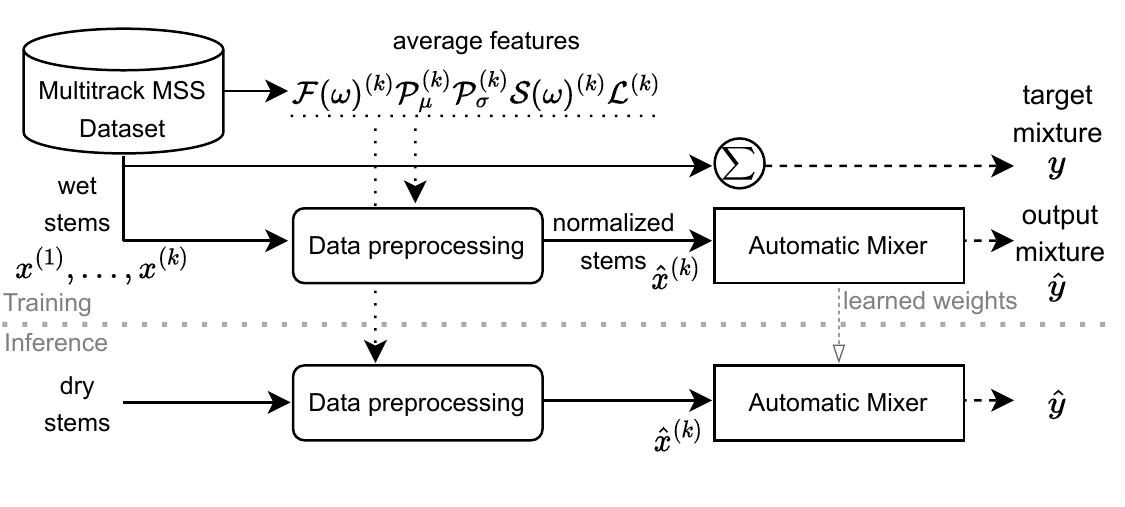}}
\caption{Our method uses data preprocessing by computing average features related to audio effects on an out-of-domain dataset (MSS data). This allows normalization of wet stems and supervised training of an automatic mixer. At inference, the same preprocessing is applied to dry data.} 
\vspace{-1.25ex}
\label{fig:automixer}
\end{figure}

Music mixing is a highly cross-adaptive transformation since the processing of an individual track depends on the content of all tracks involved. Apart from artistic considerations, it typically tries to solve the problem of unmasking by manipulating the dynamics, spatialisation, timbre or pitch of multitrack recordings~\cite{IMPbook19}. This manipulation is achieved through a set of linear and nonlinear effects, which generally can be classified into five different classes: \textit{gain, equalization (EQ), panning, dynamic range compression (DRC) and artificial reverberation}~\cite{pestana2014intelligent}. 

Several methods have been investigated to automatize this task \cite{tenyearsautomix}. For example, rule-based systems~\cite{bocko2010automatic, de2013knowledge, everardo17towards}, cross-adaptive audio mixing effects~\cite{reiss2011intelligent, pestana2013automatic} and data-driven methods~\cite{martinez2021deep,steinmetz2020mixing}. 
A black-box approach is introduced in~\cite{martinez2021deep}, where a \textit{Wave-U-Net} is trained to perform automatic mixing of drums. Conversely, neural proxies of audio effects are used as a fixed signal processing path within a deep neural framework to perform automatic mixing of songs~\cite{steinmetz2020mixing}. However, the lack of dry or unprocessed multitrack data has limited the performance of these deep learning approaches. Thus an unified approach has yet to be found that achieves results close to or superior to the quality of professional human-made mixes~\cite{moffat19approaches, wilmering20history}.
It has been hypothesized that the bottleneck of performance can be resolved with a large enough dataset~\cite{steinmetz2020mixing}. Nevertheless, collecting data is difficult, as it is unusual for musicians and record labels to provide multitrack dry recordings. 

In this work, we consider the use of out-of-domain data in conjunction with a supervised deep learning approach to close such performance gap. To achieve this, we propose a novel method that performs a data normalization or augmentation procedure on each of the audio effect classes. Figure~\ref{fig:automixer} depicts our method.
We train deep neural networks to perform automatic loudness, EQ, panning, DRC and reverberation music mixing. Thus we present 1) a data preprocessing step that allows training with out-of-domain data, 2) a new deep learning architecture, 3) an exploration of stereo-invariant loss functions, 4) a design of a perceptual listening test targeting highly skilled professionals, and 5) listening test results showing that our approach is indistinguishable from professional human-made mixes. Audio samples and code can be found at \url{https://marco-martinez-sony.github.io/FxNorm-automix/}.

\section{Method}\label{sec:method}

\subsection{Effect normalization and augmentation}

One of the main challenges of deep learning models for automatic mixing is the lack of multitrack dry data. Since collecting a large dry multitrack dataset is inherently difficult, we believe it is possible to reuse existing datasets, such as music source separation (MSS) data. Music source separation has been heavily researched in the last decade~\cite{mitsufuji2021music} and a significant effort has been put into collecting training data, such as the well-known MUSDB18 dataset~\cite{musdb18}. However, direct application of such data is infeasible, since the mixture is a summation of the wet stems, that is, mixing effects have already been applied. 


Several methods have been shown to be capable of reverse engineering the effect parameters~\cite{barchiesi2010reverse,moffat19gainDrum,colonel2021reverse}, however such approaches require the original dry multitrack recordings and target mix. There are also data-driven methods for removing effects~\cite{imort2022removing,koo2021reverb}, although they only apply to specific effects and are prone to adding sound artifacts.


Instead of removing audio effects, we propose to normalize each stem based on audio features related to each class of audio effects. In this way, different data features are scaled or transformed to make an equal contribution across the dataset~\cite{SINGH2020105524}. We propose normalization schemes for loudness, EQ, panning, DRC and reverberation, thus ensuring that all stems have been normalized to the same range of audio features. During training, we expect the models to learn how to undo or denormalize the input stems and thus approximate the original mix. At inference, we also normalize the real multitrack dry data thus allowing the model to perform automatic music mixing. This Section introduces how each effect is normalized and Section~\ref{sec:datapre} the full data preprocessing procedure.






\textbf{Loudness}\textemdash To normalize loudness, we independently compute the average loudness $\mathcal{L}^{(k)}$ for the stem type $k$ as 
\vspace{-.5ex}
\begin{equation}
\mathcal{L}^{(k)} = \frac{1}{N} \sum_{i=1}^{N} \text{LUFS}(x^{(k)}_{i}), 
\label{eq:loudness_avg}
\end{equation}

where LUFS is the integrated loudness level in dBFS in accordance to~\cite{itu2011itu}, $x$ is the $i$th stem waveform of $k$ type, e.g. $k$=vocals, and $N$ is the total number of available songs. Then, based on $\mathcal{L}^{(k)}$, each stem is loudness normalized using~\cite{steinmetz2021pyloudnorm}.

\textbf{Equalization}\textemdash 
EQ normalization is based on the average frequency magnitude spectrum $\mathcal{F}^{(k)}$ as
\vspace{-.5ex}
\begin{equation}
\mathcal{F}(\omega)^{(k)} = \frac{1}{N} \sum_{i=1}^{N} {\Gamma}(\omega)_{i}^{(k)}, \quad
{\Gamma}(\omega)^{(k)} = \frac{1}{M} \sum_{j=1}^{M} X(\omega)^{(k)}_{j}, 
\label{eq:freqeuncy_avg}
\end{equation}

where $\Gamma^{(k)}$ corresponds to the individual stem mean frequency magnitude, $\omega$ is the frequency index, $X^{(k)}$ is the magnitude of the Short-Time Fourier Transform (STFT) of each $x^{(k)}$ and $M$ its total number of frames.  

We then proceed to normalize each stem by performing EQ matching with respect to $\mathcal{F}^{(k)}$. EQ matching typically involves finding the optimal filter settings by designing and applying a filter based on the difference between the target and input frequency magnitudes \cite{germain2016equalization}.  
The differential frequency magnitude $\mathcal{F}(\omega)^{(k)}_{\text{diff}}$ is computed as 
\vspace{-.5ex}
\begin{equation}
\mathcal{F}(\omega)^{(k)}_{\text{diff}} = 10^{\log_{10} (\mathcal{F}(\omega)^{(k)}) - \log_{10} (\Gamma(\omega)^{(k)})}.
\label{eq:diff_eq}
\end{equation}

${\mathcal{F}(\omega)^{(k)}_{\text{diff}}}$ is further smoothed with a Savitzky-Golay filter \cite{schafer2011savitzky}. An FIR filter is designed using the window method \cite{gaydecki2004foundations}, and applied using forward-backward filtering \cite{smith2007introduction} to have a zero-phase response. Since forward-back filtering squares the magnitude response, we apply a square root to ${\mathcal{F}(\omega)^{(k)}_{\text{diff}}}$ before designing the FIR filter. 


\textbf{Panning}\textemdash 
The panning normalization is based on the Stereo Panning Spectrum~\cite{tzanetakis2007stereo,avendano2003frequency}, where we use the average panning as a reference then re-pan accordingly.

We compute the left and right channel similarity measure $\Psi(\omega)$, to approximate the panning gains assigned during mixing. We focus only in amplitude panning thus we calculate $\Psi(\omega)$ using only frequency magnitude as
\vspace{-.5ex}
\begin{equation}
\Psi(\omega) = 2 \frac{X(\omega)_{{\mathrm{L}}}X(\omega)_{{\mathrm{R}}}}{X(\omega)_{{\mathrm{L}}}^{2}+X(\omega)_{{\mathrm{R}}}^{2}}, 
\label{eq:psi}
\end{equation}

where $X(\omega)_{{\mathrm{L}}}$ and $X(\omega)_{{\mathrm{R}}}$ correspond to the left and right channel STFT magnitudes. $\Psi(\omega)$ denotes whether a frequency bin $\omega_{0}$ is panned to the center ($\Psi(\omega)_{\omega_{0}}$=1), or whether is panned to either side (0$\leq\Psi(\omega)_{\omega_{0}}<$1). The stereo side can be determined with the difference $\Delta(\omega)$ between partial similarities $\Psi(\omega)_{\mathrm{L}}$ and $\Psi(\omega)_{\mathrm{R}}$ as

\vspace{-2.5ex}
\begin{equation}
\hspace*{-.5ex}\Psi(\omega)_{\mathrm{L}} = \frac{X(\omega)_{{\mathrm{L}}}X(\omega)_{{\mathrm{R}}}}{X(\omega)_{{\mathrm{L}}}^{2}},  \Psi(\omega)_{\mathrm{R}}=\frac{X(\omega)_{\mathrm{L}}X(\omega)_{{\mathrm{R}}}}{X(\omega)_{{\mathrm{R}}}^{2}},
\label{eq:side}
\end{equation}

thus $\Delta(\omega)=\Psi(\omega)_{\mathrm{L}}-\Psi(\omega)_{\mathrm{R}}$, where $\Delta(\omega)$>0 and $\Delta(\omega)$<0 correspond to signals panned left or right, respectively. The panning gains, $\Phi(\omega)_{\mathrm{L}}$ and $\Phi(\omega)_{\mathrm{R}}$, then can be estimated by choosing a panning law to approximate the panning coefficient $\alpha(\omega)$. We empirically found that a linear panning law, $\Phi(\omega)_{\mathrm{L}}$=1-$\alpha(\omega)$ and $\Phi(\omega)_{\mathrm{R}}$=$\alpha(\omega)$, yields better approximations. Thus we approximate the panning gains based on $\Delta(\omega)$ and assuming $\Psi(\omega)$=2$\alpha(\omega)$. We compute the average similarity measure $\mathcal{S}(\omega)^{(k)}$ across all $\Psi(\omega)^{(k)}$ computed from $N$ stems and their STFT frames $M$. $\mathcal{S}(\omega)^{(k)}$ is also further smoothed with~\cite{schafer2011savitzky}.

For each individual stem, re-panning is implemented by 1) computing $\Psi(\omega)$ and $\Delta(\omega)$, 2) estimating $\Phi(\omega)_{\mathrm{L}}$ and $\Phi(\omega)_{\mathrm{R}}$, and 3) estimating the reference panning gains, $\hat{\Phi}(\omega)_{\mathrm{L}}$ and $\hat{\Phi}(\omega)_{\mathrm{R}}$, using $\mathcal{S}(\omega)^{(k)}$ and $\Delta(\omega)$. 
Finally, we calculate a gain factor based on the ratio of panning gains which we apply to $X(\omega)_{{\mathrm{L}}}$ and $X(\omega)_{{\mathrm{R}}}$ as

\vspace{-2.5ex}
\begin{equation}
\hat{X}(\omega)_{\mathrm{L}} = \frac{\hat{\Phi}(\omega)_{\mathrm{L}}}{\Phi(\omega)_{\mathrm{L}}} {X}(\omega)_{\mathrm{L}}, \hspace{.15ex} \hat{X}(\omega)_{\mathrm{R}} = \frac{\hat{\Phi}(\omega)_{\mathrm{R}}}{\Phi(\omega)_{\mathrm{R}}} {X}(\omega)_{\mathrm{R}}.
\label{eq:repan}
\end{equation}

Panning normalization is performed per frame and the normalized audio is obtained with the inverse STFT of $\hat{X}(\omega)_{\mathrm{L}}$ and $\hat{X}(\omega)_{\mathrm{R}}$ with their original channel phase.


\textbf{Dynamic Range Compression}\textemdash 
DRC usually alters the transients of the input~\cite{zolzer2002dafx}, thus we base our DRC normalization on the onset peak levels which are linked to such transient modification~\cite{bello2005tutorial}. 

We first perform onset detection based on the High Frequency Content (HFC) method~\cite{masri1996computer, aubio2019}. HFC emphasizes the energy variation that occurs in the upper part of the spectrum which is typical of onsets~\cite{brossier2006automatic}. We then select the maximum peak for each of the detected onsets. 

The average peak level $\mathcal{P}_{\mu}^{(k)}$ and peak level standard deviation $\mathcal{P}_{\sigma}^{(k)}$ are defined as ${\frac{1}{N}\sum_{i}^{N}}{\mu}^{(k)}_{i}$ and ${\frac{1}{N} \sum_{i}^{N}{\sigma}^{(k)}_{i}}$, respectively. Where $\mu^{(k)}$ and $\sigma^{(k)}$ are the channel mean peak level and standard deviation in dB, which for a robust estimate are calculated from the top 75th percentile of detected peaks.







DRC normalization consists in upper bounding the peak levels of the audio. Thus, if $\mu^{(k)}$ > ($\mathcal{P}_{\mu}^{(k)}$ + $\mathcal{P}_{\sigma}^{(k)}$), we apply a compressor to the input audio by performing an incremental grid search of the \textit{ratio} and \textit{threshold} parameters until $\mu^{(k)}$ < ($\mathcal{P}_{\mu}^{(k)}$ + $\mathcal{P}_{\sigma}^{(k)}$). \textit{Threshold} is the level above which compression starts, and \textit{ratio} determines the amount of compression \cite{zolzer2002dafx}. This is done with fixed \textit{attack} and \textit{release} values, i.e. the start and stop timing settings. 

\begin{figure*}[t]
\centering
\centerline{\includegraphics[width=\textwidth]{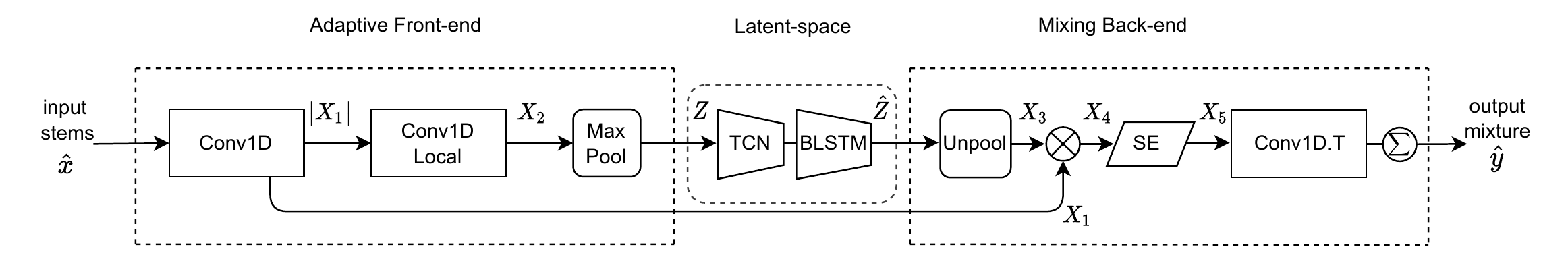}}
\caption{Block diagram of the proposed model}
\label{fig:model}
\end{figure*}

\textbf{Artificial Reverberation}\textemdash 
Due to the inherent characteristics of reverberation~\cite{zolzer2002dafx}, an attempt to similarly normalize this effect is not trivial. Blind estimation of reverberation features, such as reverberation time (RT) or direct-to-reverberant ratio, is an open research area in itself, and considering that most of the research has been done for speech signals~\cite{eaton2016estimation, lollmann2010improved, duangpummet2021blind}, its application to music is beyond the scope of this paper.

We propose instead a data augmentation approach where we stochastically add reverberation to already reverberated stems. Reverberation is added using a mixing method called the 'Abbey Road reverb trick'~\cite{white2020processing}, where a chain of EQ and reverb effects is applied as a send effect, i.e. a copy of the input is processed and added to the input. Naively adding reverberation directly to the input audio has the potential to clutter the low and low-mid frequencies. 
In this manner, the process of learning how much reverberation is required for a given mix is carried out by the network by learning to filter out the additional reverberation that is present in the input stems.

\subsection{Data preprocessing}\label{sec:datapre}

Since most mixing effects are interrelated, applying the normalization methods separately yields inaccurate results, e.g. EQ normalization modifies the dynamics of the audio, thus modifying DRC features. To minimize this, we perform effect preprocessing progressively and based in the following order: EQ, DRC, panning and loudness. Since EQ and DRC normalization are done on a per channel basis, panning normalization is applied after the above to avoid further modification of the stereo features.

Thus, for each stem type $k$, average feature computation corresponds to 1)~calculate $\mathcal{F}(\omega)^{(k)}$ and EQ normalize, 2)~calculate $\mathcal{P}^{(k)}_{\mu}$ and $\mathcal{P}^{(k)}_{\sigma}$ and DRC normalize, 3)~calculate $\mathcal{S}(\omega)^{(k)}$ and panning normalize, and 4)~calculate $\mathcal{L}^{(k)}$ and loudness normalize.

We apply our reverberation preprocessing method after all the average features have been calculated.
Considering that this procedure is based on a stochastic data augmentation procedure, for consistency we treat the added reverberation as noise for the normalization pipeline. The final data preprocessing method corresponds to applying reverberation augmentation followed by EQ, DRC, panning, and loudness normalization methods.

\subsection{Architecture}

We propose a new architecture based on \cite{martinez2020deep} and  \cite{luo2019conv}. It operates in the time-domain and processes raw waveforms in a frame-wise manner. The model can be divided into three parts: adaptive front-end, latent-space mixer and synthesis back-end. The model is depicted in Figure \ref{fig:model} and its architecture is summarized in Table \ref{table:architecture_ours}.

The \textbf{adaptive front-end} is exactly the same as in \cite{martinez2020deep}, with the only difference that now the input $\hat{x}$ corresponds to $K$ stereo tracks of length $A$. In the front-end, time-domain convolutions are applied to the input audio in order to learn a latent representation $\boldsymbol{Z}$ and a filter bank which output feature map $\boldsymbol{X}_{1}$ corresponds to a frequency band decomposition of $\hat{x}$. The \textbf{latent-space mixer} corresponds to the temporal dilated convolutions (TCN) separator block of \cite{luo2019conv}, and to improve long-term dependencies learning, the TCN is followed by stacked Bidirectional Long Short-Term Memory (BLSTM) layers. Contrary to \cite{luo2019conv}, the objective of the mixer is not to learn a mask for source separation, but rather to learn a mixing mask $\hat{\boldsymbol{Z}}$. 



The \textbf{synthesis back-end} uses $\hat{\boldsymbol{Z}}$ to modify $\boldsymbol{X}_{1}$ based on the given mixing task and its design is motivated by~\cite{martinez2020deep}. It upsamples the mixing mask using nearest neighbor interpolation and via an element-wise multiplication applies it to each filter-bank channel source of $\boldsymbol{X}_{1}$ at each time step. We hypothesize that this frequency-based transformation is akin to the model learning and applying a dynamic equalization effect~\cite{valimaki2016all} while also filtering out the extra reverberant content of the normalized input stems. 

The resulting feature map $\boldsymbol{X}_{4}$ is further modified via a Squeeze-and-Excitation (SE) block \cite{hu2018squeeze} implemented as shown in~\cite{martinez2020deep}. The SE layers scale the channel-wise information of $\boldsymbol{X}_{4}$ by applying an adaptive gain $\boldsymbol{s}_{c}$, and consequently, learning a loudness gain and panning transformation for each filter bank channel. The modified frequency decomposition $\boldsymbol{X}_{5}$ is then reconstructed using a non-trainable transposed convolution as shown in~\cite{martinez2020deep}. Finally, the resulting $2K$ stereo tracks are channel-wise summed into a stereo output and a hyperbolic tangent is used to avoid clipping. All convolutions use a stride of 1 to avoid ringing and filtering artifacts \cite{pons2021upsampling}. BLSTMs and SE layers are applied to the filter dimension.

\begin{table}[t]
\caption{Summarized architecture of the proposed model. }
\centering
\resizebox{.54\columnwidth}{!}{
\renewcommand{\arraystretch}{1.0}
  \begin{tabular}{cccc}
  \toprule 
  Layer & Output shape & Output \\ [1ex]
  \midrule\midrule
  Input stems & ($2K$, $A$)  & $\boldsymbol{\hat{x}}$\\
  Conv1D  & ($N$, $A$)  & $\boldsymbol{X}_{1}$\\
  Conv1D-Local  & ($N$, $A$)  & $\boldsymbol{X}_{2}$ \\
  MaxPooling & ($N$, $A/64$)  & $\boldsymbol{Z}$\\
  \midrule
  TCN & ($N$, $A/64$)  &. \\
  BLSTM & ($A/64$, $N$)  & $\boldsymbol{\hat{Z}}$\\
  \midrule
  Unpooling & ($N$, $A$)  & $\boldsymbol{X}_{3}$\\
  $\boldsymbol{X}_{3}\times\boldsymbol{X}_{1}$  & ($N$, $A$)  & $\boldsymbol{X}_{4}$\\
  \hdashline
  SE (Abs) & ($A$, $N$) &. \\
  SE (Global Avg) & (1, $N$)  &.\\
  SE (FC) & (1, $16N$)  &. \\
  SE (FC) & (1, $N$)  & $\boldsymbol{s}_{c}$\\
  \hdashline
  $\boldsymbol{X}_{4}\times s_{c}$ & ($N$, $A$)  & $\boldsymbol{{X}}_{5}$\\
  Conv1D.T & ($2K$, $A$)  &.\\
  Summation & ($2$, $A$)  & $\hat{y}$\\
  \bottomrule   
  \end{tabular}
  }
\label{table:architecture_ours}
\end{table}

\subsection{Loss function}

Based on the stereo-invariant loss function introduced by~\cite{steinmetz2020mixing}, we explore two variations of such loss. First, due to the perceptual nature of the mixing task and motivated by \cite{wright2020perceptual}, we apply A-weighting pre-emphasis and low-pass FIR filters ($\rho$) to the target and output audio frames $y$ and $\hat{y}$, respectively. Then we compute the sum and difference signals $y_{\text{sum}}$=$\rho(y_{\mathrm{L}})$+$\rho(y_{\mathrm{R}})$ and $y_{\text{diff}}$=$\rho(y_{\mathrm{L}})$-$\rho(y_{\mathrm{R}})$. 

The first loss follows closely \cite{steinmetz2020mixing} and is based on the Spectral Convergence (SC), magnitude-normalized Frobenius norm, and the L1-norm spectral log-magnitude (L1Log) as
\vspace{-.5ex}
\begin{equation}
\begin{split}
L_{\text{a}} = l_{\text{SC}}(Y_{\text{sum}},\hat{Y}_{\text{sum}}) + l_{\text{L1Log}}(Y_{\text{sum}},\hat{Y}_{\text{sum}})\\  + l_{\text{SC}}(Y_{\text{diff}},\hat{Y}_{\text{diff}})+l_{\text{L1Log}}(Y_{\text{diff}},\hat{Y}_{\text{diff}}),
\end{split}
\label{eq:lossa}
\end{equation}

where $Y$ and $\hat{Y}$ are the respective 4096-Fast Fourier Transform (FFT) magnitude with a 25\% hop size.

The second loss replaces the SC loss component with a less penalizing normalization such as the widely used L2-norm on the spectral magnitude (L2), defined as

\vspace{-3.5ex}
\begin{equation}
\begin{split}
L_{\text{b}} = l_{\text{L2}}(Y_{\text{sum}},\hat{Y}_{\text{sum}}) + l_{\text{L1Log}}(Y_{\text{sum}},\hat{Y}_{\text{sum}})\\  + l_{\text{L2}}(Y_{\text{diff}},\hat{Y}_{\text{diff}})+l_{\text{L1Log}}(Y_{\text{diff}},\hat{Y}_{\text{diff}}).
\end{split}
\label{eq:lossb}
\end{equation}

We empirically found the perceptual-based pre-emphasis filter as vital when modeling a highly perceptual task such as music mixing. We also found an easier convergence during training when using a single frame-size loss rather than a multi-resolution magnitude loss.

\section{Experiments}\label{sec:experiments}

\subsection{Dataset}

We first conduct experiments with a small dataset (S) which corresponds to the MUSDB18 dataset \cite{musdb18}. This dataset consists of wet stems for \textit{vocals}, \textit{drums}, \textit{bass} and \textit{other} ($K$=4), where the \textit{mixture} is the summation of such stems. There are a total of 150 songs, of which 86 are used for training and 14 and 50 for validation and testing purposes, respectively. 

We also use a private large dataset (L) for music separation which corresponds to 1,505 extra multitrack songs created in the same manner as MUSDB18. We incorporate the MUSDB18 training and validation sets into the large dataset, thus obtaining a total of 1,455+86=1,541 training songs and 50+14=64 validation songs. In both datasets, most of the songs correspond to mainstream western music, with rock and pop as the predominant genres.

To validate our method, we use a private set of 18 dry multitrack songs, where all songs have been produced by different musicians and mixing engineers. From this data we grouped the respective dry stems without applying any effect in the process.




\subsection{Dataset preprocessing}
\label{sec:preprocessing}
We apply our preprocessing methods to both datasets independently. EQ preprocessing is computed with a 65,536 point STFT with hop size of 25\% and a FIR filter of 1,001 taps. In order to avoid clipping, all audio stem channels are loudness normalized to -30 dBFS prior to the EQ feature computation and normalization. 

For DRC preprocessing, to determine the timing settings of the compressor used during the normalization, we averaged the values found in mixing engineering best practices~\cite{pestana2013automatic,de2017towards}. \textit{Attack} values correspond to 7.5, 10, 10 and 15 ms and \textit{release} values to 400, 180, 500 and 666 ms, for \textit{vocals}, \textit{drums}, \textit{bass} and \textit{other}, respectively. For the incremental grid search we use \textit{ratio} and \textit{threshold} values within \{4, 20\} and \{-40, -10\} db respectively. For the HFC computation we use 128 mel bands of all stems with the exception of bass where we found better onset detection with 16 mel bands. Prior to the DRC preprocessing, all channels are peak normalized to -10 dB.

Panning preprocessing is done with a 2,048 point STFT with hop size of 50\%. For all stems, frequency bins are re-panned until 16 kHz, with the exception of \textit{drums}, which is re-panned until 16 kHz in order to avoid artifacts.

Reverberation augmentation is done by applying uniform random sampling on a set of 130 different impulse responses (IR) whose RT is within \{2, 4\} seconds (\textit{s}). The EQ used prior to the reverberation corresponds to a low-shelf and high-shelf with a fixed gain of -30 dB whose cutoff frequency is uniformly sampled within \{500, 700\} Hz and \{7, 10\} kHz, respectively. At inference, to simulate the reverberant characteristics of the training data, before applying the aforementioned augmentation, a "pre-reverb" is added in the same manner but from a different set of 400 IRs whose RT is within \{1, 1.5\}-\textit{s}. A shorter RT is chosen, as high reverberation levels has been shown to have a more detrimental effect on subjective preference than low levels \cite{de2017perceptual}. Reverberation augmentation is applied only to \textit{vocals} and \textit{other} stems.
Reverberation and DRC effects are implemented using the Python package from~\cite{steinmetz2020mixing}.

\subsection{Hyperparameters}

The convolutional layers on the front-end have $N$=128 filters of size 64 and 128 respectively, and a 64-point window for max-pooling. In the mixer, the bottleneck layer has 256 channels, the skip connection paths have 64 channels, and the convolutional blocks have 128 channels of kernel size 3. We use 4 stacks of 6 convolutional layers with a dilation factor of 1,2,...,32. We stacked 3 BLSTMs with a hidden feature map of 64 channels. The FC layers in the SE block have 2048 (16$N$) and 128 channels respectively.

The input consists of 2$K$ channels, each channel consisting of $A$=10-\textit{s} audio frames at 44.1 kHz. The receptive field of the mixer is 505 samples, and taking into account the pooling operation, the overall network has a receptive field of 32,446 samples. Thus, the loss function is only calculated on a 8.52-\textit{s} frame centered in the middle of the 10-\textit{s} input. The network has in total 2.7M parameters.
As a baseline, we use the modified \textit{Wave-U-Net} (WUN) introduced by \cite{martinez2021deep}. The same settings are used as in the original work which yields a network of 2.5M parameters. 

\subsection{Training}

Regarding on-the-fly data augmentation, we apply a randomization of the order of the stereo channels, this is done consistently across the stems and the target mix. For the proposed network we use the pretraining from~\cite{martinez2020deep}. We train both models using a batch size of 4, an initial learning rate $\beta$=0.001, and the following learning rate schedule; $\beta$ for 300 epochs, $\beta$/3 for 100 epochs, $\beta$/10 for 100 epochs, $\beta$/30 for 50 epochs, $\beta$/100 for 50 epochs and $\beta$/1000 for 25 epochs. An epoch consists of 1600 batches, L2 norm of the gradient is clipped by 0.2 and we use 10$^{-7}$ for weight decay regularization. We select the model with the lowest validation loss.


\section{Results \& Analysis}\label{sec:results}

We trained both networks with the preprocessed datasets S and L and the loss functions $L_{a}$ and $L_{b}$, which yielded Ours-S-L$_{a}$, Ours-S-L$_{b}$, Ours-L-L$_{a}$ and Ours-L-L$_{b}$ for our proposed network, and WUN-S-L$_{b}$ and WUN-L-L$_{ b}$ for Wave-U-Net. Convergence during training for WUN with $L_{a}$ was unsuccessful, therefore it is excluded from the results.

\subsection{Quantitative evaluation}

To measure how close the output mixes are to the reference mixes, we use the following audio features, spectral: centroid, bandwidth, contrast, flatness, and roll-off \cite{peeters2004large}; panning: the Panning Root Mean Square (RMS)~\cite{tzanetakis2007stereo}; dynamic: RMS level, dynamic spread and crest factor \cite{ma2015intelligent}; and LUFS loudness level~\cite{itu2011itu}. All features are computed using a running mean of 0.5-\textit{s}~\cite{tzanetakis2007stereo}. Table \ref{tab:objective_metrics_spec} shows the results for the dry test set and the MUSDB18 test set. As expected, the overall error values are larger for the dry test set. Also, the loudness, dynamics, and panning values show a closer match to the reference mix when compared to the Normalized mix, which is the sum of input stems after data preprocessing. Spectral values do not match in the same way, which could indicate that the generated mixes deviate in terms of EQ and reverberation.

\begin{figure*}[!t]
\centering
\centerline{\includegraphics[trim=0cm 0.275cm 0cm 0cm, clip, width=.9\textwidth]{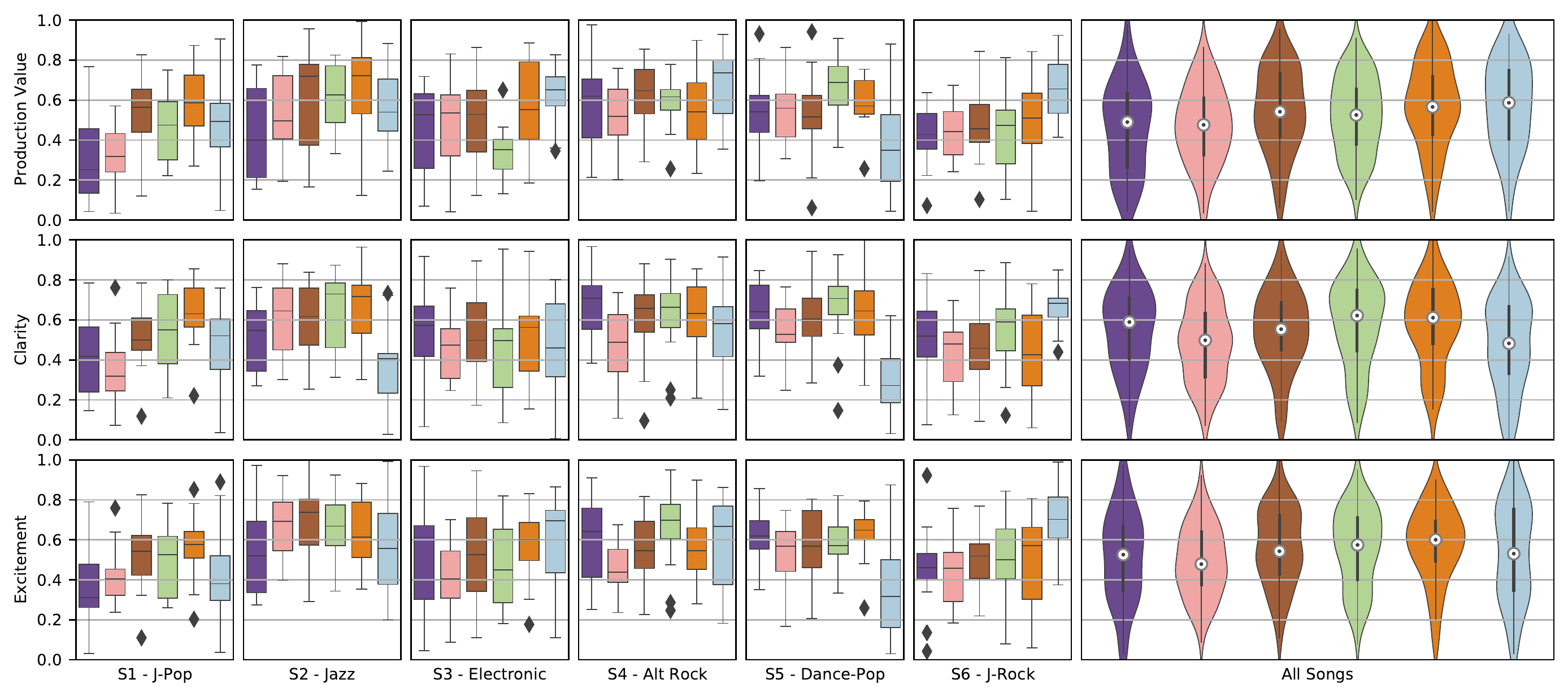}}
\begin{minipage}[b]{.99\textwidth}
  \centering
  \centerline{\includegraphics[trim=0 0.3cm 0 0, clip, width=0.55\columnwidth]{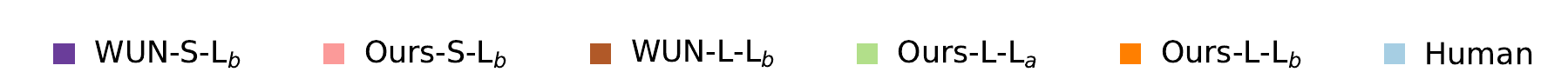}}
\end{minipage}
\caption{Listening test boxplots and violin plots for all individual six songs (S1 to S6) and all songs, respectively.}
\label{fig:boxplot}
\end{figure*}

\begin{table}[!t]

\resizebox{1.0\columnwidth}{!}{
\begin{tabular}{p{2cm}p{1cm}p{1cm}p{1cm}p{1cm}|p{1cm}p{1cm}p{1cm}p{1cm}}

\toprule
 & \multicolumn{4}{c}{dry test set} & \multicolumn{4}{c}{MUSDB18 test set} \\
model & spectral & panning & dynamic & loudness & spectral & panning & dynamic & loudness  \\
\midrule\midrule
Normalized        & \underline{{0.435}}          & {0.593}           & {0.168} & {0.633}  & \underline{{0.256}}           & {0.783}           & {0.109} & {0.643}  \\
WUN-S-L$_{b}$      & {0.527}           & {0.215}           & {0.082} & {0.094}    & \textbf{{0.250}}  & {0.250}           & \underline{{0.078}} & {0.097}   \\
Ours-S-L$_{a}$     & {0.547}           & {0.201}           & {0.062} & \underline{{0.056}}  & {0.299}           & \underline{{0.191}}           & {0.086} & {0.095}  \\
Ours-S-L$_{b}$     & \textbf{{0.427}}  & {0.207}           & {0.063} & {0.061}    & {0.276}           & {0.212}           & {0.085} & {0.084}   \\
WUN-L-L$_{b}$      & {0.551}           & \underline{{0.182}}         & {0.066} & \textbf{{0.054}}   & {0.279}           & {0.195}          & \textbf{{0.074}} & \textbf{{0.072}}   \\
Ours-L-L$_{a}$     & {0.590}           & {0.191}           & \textbf{{0.055}} & {0.091}   & {0.312}           & {0.593}           & {0.168} & {0.105}   \\
Ours-L-L$_{b}$     & {0.519}           & \textbf{{0.170}} & \underline{{0.056}} & {0.061}   & {0.276}           & \textbf{{0.160}} & {0.084} & \underline{{0.079}}   \\
\bottomrule
\end{tabular}
}
\caption{Objective metrics correspond to the average mean absolute percentage error by feature subgroup.}
\label{tab:objective_metrics_spec}
\end{table}

\subsection{Listening Test}

We designed a the listening test using the Web Audio Evaluation Tool~\cite{jillings15WebAudio} and the APE interface~\cite{de2014ape}. The test is intended for professional mixing engineers only and we ask participants to rate the mixes based on Production Value, Clarity and Excitement as shown in~\cite{pestana2014cross}. Perceptual tests for mixing systems \cite{steinmetz2020mixing, martinez2021deep} often differ from MUSHRA \cite{recommendation2001bs} tests, as the reference sample is omitted in order to encourage a direct comparison between mixes. However, based on feedback from pilot tests, we decided to include the 4 dry stems in the test as references. In this way, the ratings may reflect the quality of transformation on each stem as applied by the mixing systems.



Fourteen participants with an average mixing engineering experience of 11.6 years took part in the test. In total there were six different songs and for each song six different mixes were presented. Each mix was 25-\textit{s} taken from the chorus-to-verse transitions of the full mixes. The six mixes correspond to the Ours-S-L$_{b}$, Ours-L-L$_{a}$, Ours-L-L$_{b}$, WUN-S-L$_{b}$ and WUN-L-L$_{b}$ models plus a professional Human mix. Ours-S-L$_{a}$ was omitted from the test to allow more songs to be tested while avoiding listening fatigue~\cite{schatz2012impact}. A low-anchor was not used as it has been shown to compress the other ratings at the higher end~\cite{IMPbook19}. All mixes were loudness normalized to -$23$ dBFS~\cite{ebu2011loudness}.

Figure \ref{fig:boxplot} shows the results of the listening test and Table \ref{tab:signif} shows the pairwise comparisons of mixes. For Production Value, there is no statistically significant difference between the Human mixtures and the models that were trained on the large dataset. For Clarity, Human mixes are rated lower than Ours-L-L$_{a}$, Ours-L-L$_{b}$ with a p-value of 0.008 and 0.013, respectively, indicating that our models generate mixtures with less masking. For Excitement, the null hypothesis is accepted, and therefore Human mixes are considered no different than any other mixture system. Although Ours-L-L$_{b}$ was among the best among all criteria, the null hypothesis is also accepted when compared to WUN-L-L$_{b}$. Thus, a further analysis is needed among both networks, e.g. in terms of long-term learned dependencies and ability to generate full-length coherent mixes.

In general, models trained with large datasets received the highest scores, which could confirm that lack of data has been the bottleneck of data-driven mixing systems. However, for Clarity and Excitement, the mixes by WUN-S-L$_{b}$ are not considered different from the rest, which might indicate that our data preprocessing method is both effective for small and large datasets.



It should be noted that, in general, none of the mixes is consistently considered as very good. This relates with other findings, where even commercial mixes made by renown engineers are not rated as high~\cite{de2017towards, IMPbook19}. This opens up a new research direction for evaluating mixing systems, as this type of test is too difficult for inexperienced participants, and when experienced participants do participate, they tend not to rate any mix as very good. 

For individual songs, Human mixes have lower rates for Jazz and Dance-Pop. For the latter, low ratings on all criteria may be due to highly compressed drums. However, the high ratings in terms of Excitement and Production Value for the Human J-Rock mix might be due to hard-panned guitars. In contrast, the models are conservative when it comes to panning and are unlikely to hard-pan sources, however a further analysis is required. Furthermore, although we show the models successfully mixing multiple genres, an in-depth analysis of the genre distribution of the dataset with ratings by genre is needed.

\begin{table}[t]
\centering
\hspace*{-.5cm}
\resizebox{.99\columnwidth}{!}{
\begin{tabular}{m{0.15\textwidth}|ccccccc}
\toprule
Production Value & WUN-S-L$_{b}$ & Ours-S-L$_{b}$ & WUN-L-L$_{b}$ & Ours-L-L$_{a}$ & Ours-L-L$_{b}$ & Human  \\
\midrule\midrule
WUN-S-L$_{b}$  &            $\cdot$ &         o &     $\ubar{\star}$ &         o &        $\ubar{**}$ &   $\ubar{\star}$  \\
Ours-S-L$_{b}$ &                  o &   $\cdot$ &                     o &       o &             $\ubar{**}$ &   $\ubar{\star}$  \\
WUN-L-L$_{b}$  & $\obar{\star}$ &         o &               $\cdot$ &       o &                          o &            o  \\
Ours-L-L$_{a}$ &                 o &         o &                     o &   $\cdot$ &                        o &         o  \\
Ours-L-L$_{b}$ &  $\obar{**}$ & $\obar{**}$ &                o &         o &                    $\cdot$ &        o  \\
Human       & $\obar{\star}$ & $\obar{\star}$ &               o &         o &                         o &        $\cdot$  \\
\bottomrule
\end{tabular}}

\centering
\hspace*{-.5cm}
\resizebox{.99\columnwidth}{!}{
\begin{tabular}{m{0.15\textwidth}|ccccccc}
Clarity & WUN-S-L$_{b}$ & Ours-S-L$_{b}$ & WUN-L-L$_{b}$ & Ours-L-L$_{a}$ & Ours-L-L$_{b}$ & Human  \\
\midrule\midrule
WUN-S-L$_{b}$  &   $\cdot$ &                      o &       o &                      o &                   o &          o  \\
Ours-S-L$_{b}$ &         o &                $\cdot$ &       o &         $\ubar{*}$ &     $\ubar{**}$ &         o  \\
WUN-L-L$_{b}$  &        o &                      o &   $\cdot$ &                      o &                   o &          o  \\
Ours-L-L$_{a}$ &         o &        $\obar{*}$ &       o &                 $\cdot$ &                    o &     $\obar{*}$  \\
Ours-L-L$_{b}$ &         o &        $\obar{**}$ &       o &                       o &               $\cdot$ &   $\obar{\star}$  \\
Human       &         o &                       o &        o &       $\ubar{*}$ &    $\ubar{\star}$ &  $\cdot$  \\
\bottomrule
\end{tabular}}

\centering
\hspace*{-.5cm}
\resizebox{.99\columnwidth}{!}{
\begin{tabular}{m{0.15\textwidth}|ccccccc}
Excitement & WUN-S-L$_{b}$ & Ours-S-L$_{b}$ & WUN-L-L$_{b}$ & Ours-L-L$_{a}$ & Ours-L-L$_{b}$ & Human  \\
\midrule\midrule
WUN-S-L$_{b}$  &   $\cdot$ &                 o &                   o &                     o &                 o &           o     \\
Ours-S-L$_{b}$ &         o &           $\cdot$ &      $\ubar{*}$ &   $\ubar{\star}$ &  $\ubar{*}$ &       o  \\
WUN-L-L$_{b}$  &        o &    $\obar{*}$ &              $\cdot$ &                    o &                 o &               o  \\
Ours-L-L$_{a}$ &         o &    $\obar{\star}$ &                 o &               $\cdot$ &                o &             o  \\
Ours-L-L$_{b}$ &         o &       $\obar{*}$ &                  o &                     o &          $\cdot$ &        o  \\
Human       &         o &                    o &                  o &                    o &                o &        $\cdot$  \\
\bottomrule
\end{tabular}}

\caption{Post hoc Mann-Whitney test results of pairwise comparison with Bonferroni Correction. o > 0.05, $\star$ < 0.05, $*$ < 0.01, $**$ < 0.001. E.g. when y-axis is compared to x-axis, $\bar{*}$ or $\ubar{*}$ indicate y-axis is significantly better or worse than x-axis for a p-value < 0.01, respectively. }

\label{tab:signif}
\end{table}

\section{Conclusion}\label{sec:conclusion}

We present a novel data preprocessing approach that allows us to train deep learning networks with existing wet or processed multitrack data by reusing them to perform an automatic mixing task. During inference, we apply the same data preprocesing to dry multitrack data and we tested the generated mixes via objective and subjective tests. We introduced a new deep learning architecture designed for the proposed task, a perceptual-based loss function along with a redesigned listening test aimed at experienced mixing engineers. The results indicate that our approach successfully achieves automatic loudness, EQ, DRC, panning, and reverberation music mixing. Resulting mixes compared to professional mixes scored higher in terms of Clarity and are indistinguishable in terms of Production Value and Excitement. We believe that the proposed preprocessing can be applied to other data-driven MIR tasks.

\clearpage

\section{Acknowledgments}\label{sec:acknowledgments}

We would like to express special thanks to K. Gokan and S. Masui for their valuable comments and all the mixing engineers from Sony Music Entertainment Japan, Sony Europe and Queen Mary University of London who participated in the listening test.

\bibliography{ISMIRtemplate}

\end{document}